\documentstyle[epsfig]{article}

\newcommand{\ndt}{\noindent}

\newcommand{\ov}{\overline}

\begin{document}

\begin{flushright}
{\large {\bf DFUB 13/01}}
\end{flushright}


\begin{center}
\vspace{1cm}
{\Large {\bf MAGNETIC MONOPOLES}}

\vspace{1cm}

G. GIACOMELLI and L. PATRIZII

\vspace{5mm}

{\it Dipartimento di Fisica dell'Universit\`a di Bologna \\ and INFN,
 Sezione di Bologna\\
           Viale C. Berti Pichat 6/2, I-40127, Bologna, Italy\\
E--mail: giacomelli@bo.infn.it, patrizii@bo.infn.it}




\vskip 1.0 truecm

\vspace{5mm}

Invited paper at the NATO ARW ``Cosmic Radiations: from Astronomy to 
Particle Physics", Oujda, Morocco 21-23 March 2001

\end{center}

{\small Abstract. We discuss the experimental 
situation of
 direct searches at accelerators for Dirac magnetic monopoles, and in 
the penetrating cosmic radiation for the superheavy
 magnetic monopoles predicted by GUT theories. We also discuss the searches
for intermediate mass monopoles (which are predicted by theories in extra
dimensions), and for nuclearites and Q-balls.}

\section{Introduction}
The magnetic monopole (MM) concept may be traced back
to the origin of ma\-gne\-tism, with first accounts in 1269.
At the beginning of the 19th century there were discussions and experiments
concerning the magnetic content of matter and some speculations about the
possible existence of isolated magnetic charges.\par
 In 1931 Dirac introduced the MM in order to explain the 
quantization of the electric charge, which
 follows from the existence of at least one free magnetic 
charge \cite{dirac}. Dirac established the relationship between the basic 
elementary electric
charge $e$ and the basic magnetic charge $g$:~
$eg=n\hbar c/2$, where $n$ is an integer;
 $g_{D}=\hbar c/2e$ is called the unit Dirac charge. There was no prediction
 for the MM mass.\par 
 From 1931 many searches for  
``classical Dirac monopoles'' were carried out at every new accelerator;  
the searches were made with  relatively simple set--ups.
\par
In 1974 it was realized that the electric charge is naturally quantized 
in  unified gauge theories of the basic interactions 
 and that such unified theories
imply the existence of MMs, with  calculable properties. 
In the context of the Grand Unification Theory of strong and electroweak 
 interactions (GUT), the MMs appear at the phase transition 
corresponding to the spontaneous breaking of the unified group into subgroups,
one of which is U(1) \cite{thooft}. The  MM     
mass is related to the mass $m_{X}$ of  the carriers X of the
unified interaction, $ m_{M}\ge m_{X}/G$, 
where G is the dimensionless unified coupling constant at E $\simeq m_{X}$. 
In GUT one has 
$m_{X}\simeq 10^{14}-10^{15}$ GeV and $G\simeq 0.025$; 
consequently $m_{M}> 10^{16}-10^{17}$ GeV. 
This is an enormous
mass: MMs cannot be produced at any man--made accelerator, 
existing or conceivable. They could only be produced in the first instants of 
our  universe and they may be searched  for as relic particles in the 
penetrating cosmic radiation. \par
 Larger MM masses are expected
 if gravity is brought into the unification 
 picture and in some  SuperSymmetric theories.\par

The application of the simplest GUTs to the standard early universe 
scenario yields too many monopoles, while inflationary scenarios
lead to a very small number. Thus
 gauge theories of the unified interactions demand the existence of
MMs; however,  the prediction of the monopole mass is uncertain by 
several orders of magnitude, the magnetic charge could be anywhere 
between one and 
several Dirac units, and the expected flux could vary from a very small
value to a sizeable and observable one. \par

Intermediate mass monopoles (IMMs) could have been produced in later
 phase transitions in the early universe, in which a semisimple gauge group
yields a U(1)
factor at a lower energy scale. IMMs with masses around
$10^{7} \div 10^{13}$ GeV have been proposed \cite{lazaride,kephart}.
Superheavy MMs and IMMs are   topological point defects; an
undesirable large number of relatively light monopoles may be gotten rid of by
means of higher 
dimensional topological defects (strings, walls).\par
One of the recent interests in relatively low mass MMs is connected also
with the possibility that relativistic MMs could be the source of the
highest energy cosmic rays, with energies larger than $10^{20}$ eV 
\cite{kephart}.
Intermediate mass MMs could be accelerated to relativistic velocities
 in one coherent domain of the galactic
 magnetic field, or in the intergalactic field, or  in several astrophysical
sites,
like in the magnetic fields of Active Galactic Nuclei (AGN)
 and even of neutron stars.

The lowest mass MM is  expected to be stable, since magnetic
charge should be conserved like electric charge. Therefore, the  
MMs produced in the early universe should still exist as cosmic 
relics, whose kinetic energy has been strongly affected by their travel 
through galactic and intergalactic magnetic fields.\par
\par
 The most direct method of searching for GUT monopoles is to search for them 
 in the penetrating cosmic radiation. GUT poles should be characterized by low velocities
and relatively large energy losses. 
Instead IMMs should be relativistic and should be searched for at  
high altitude laboratories, and possibly at sea level via Cherenkov radiation.\par

In the following we shall summarize the basic properties of MMs, and of their 
 interactions
in matter. Searches for classical, GUT and intermediate mass monopoles  
 are then described. Monopole catalysis of proton decay is 
discussed
in Sect.~9. An outlook 
and conclusions are given in Sect.~12. 
We shall also briefly discuss searches for nuclearites and Q-balls.

\section{Main properties of magnetic monopoles}
The
consequences of the Dirac relation, 
$  eg = {n \hbar c/2}$, are 
 summarized  here.\par

\noindent - {\it Magnetic charge.} 
If $n$~=1 and if the basic electric charge is that of the electron, then  
the basic magnetic charge is 
$ g_D =\hbar c/ 2e=137e/2=3.29\times 10^{-8}\quad
 CGS.$

\noindent - {\it Coupling constant.} 
In analogy with the fine--structure constant, $\alpha =e^{2}/\hbar c\simeq 
1/137$, the 
 dimensionless magnetic--coupling constant is 
$ \alpha_g=g^{2}_{D}/ \hbar c \simeq 34.25.$
\par
\ndt - {\it Energy W acquired in a magnetic field  B}:~  
$  W = ng_{D} B\ell = n 20.5$ keV/G~cm.
\par\ndt 
 In a coherent galactic--length,  
  $\ell\simeq 1$ kpc, and $B\simeq 3~\mu$G, the energy gained  
by a monopole is:
 $ W_G=WB\ell\simeq 1.8\times 10^{11}$ GeV.

\par
\noindent- {\it Energy losses in matter.} 
A fast MM with magnetic charge $g_D$ and velocity $v=\beta c$ 
behaves like an equivalent electric charge 
$(Ze)_{eq}=g_D\beta$.
                  
\noindent- {\it Trapping of MMs in ferromagnetic materials.}  
MMs may be
trapped in  ferromagnetic materials by an image force, 
which  may reach the value of $\simeq 10$ eV/\AA.
\par
\noindent - {\it Mass and spatial structure of a GUT pole}
 (with $m_M\simeq 10^{17}$ GeV). It 
may be pictured as having:
(i) a core with radius $r_{c}\simeq 1/m_{X}\simeq 10^{-29}$ cm; 
(ii) a region up to $r\simeq 10^{-16}$ cm, where virtual $W^{+}$, $W^{-}$ and 
$Z^{o}$ may be present; (iii) a confinement region with $r_{conf}\simeq 1$ fm; 
(iv) a fermion--antifermion condensate region up to $r_{f}\simeq 1/m_{f}$; the 
condensate may contain 4--fermion baryon--number--violating terms;
(v) for $r\geq  3$ fm a MM behaves as a point particle  
 which generates a field $B=g/r^{2}$ (see Fig.~1) \cite{picture}.\par 
\noindent- Electrically
charged monopoles (dyons) may arise as quantum--mechanical excitations of GUT 
poles or as M--p, M-nucleus composites.\par
\noindent- The structure of an IMM would be similar to that of a GUT 
monopole, but the core would be larger (since R $\sim$ 1/$m_M$) and the outer 
cloud would not contain 4--fermion baryon--number--violating terms. \par

\begin{figure}
\begin{center}
\mbox{
        \epsfig{file=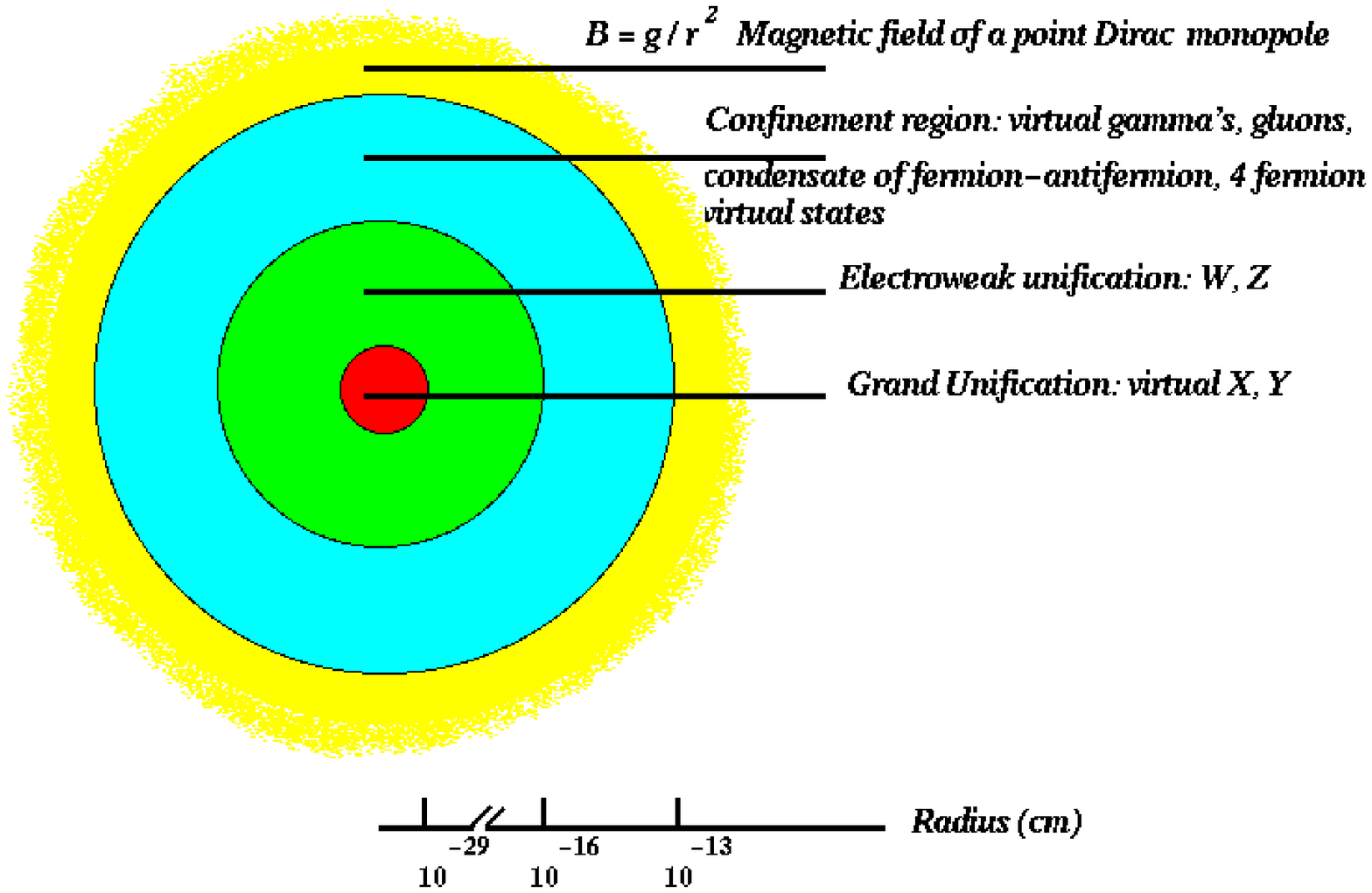,height=7.4cm}}
\end{center}
\vspace{0.5cm}
{\small Figure:~1 Structure of a GUT monopole. The various regions correspond
to: (i) Grand Unification ($r \sim 10^{-29}$ cm; inside this core one finds
virtual $X$ and $Y$ particles); (ii) electroweak unification
($r \sim 10^{-16}$ cm; inside this region one finds virtual $W^{\pm}$
and $Z^0$); (iii)
confinement region ($r \sim 10^{-13}$ cm; inside one finds virtual
$\gamma$, gluons and a condensate of fermion-antifermion pairs and 4-fermion
virtual states); (iv) for $r>$ few fm one has the field of a point
magnetic charge.}
\end{figure}

\section{Interactions of magnetic monopoles with matter}
It is important to know whether 
the quantity and quality of
energy lost by a MM in a particle detector is adequate for 
its 
detection. Classical poles and IMMs can be 
accelerated to relativistic velocities. 
  Instead GUT poles have large masses and are expected 
to have relatively low velocities, $10^{-4}<\beta<10^{-1}$, $\beta=v/c$. 
 The interaction of the MM magnetic charge with nuclear magnetic dipoles 
 could lead 
to the formation of M--nuclei bound systems. 
 This may affect the energy loss
in matter and the cross--section for MM catalysis of proton decay.
 A monopole--proton bound state may be produced via  radiative 
capture,  
$ M+p\to    (M+p)_{bound} + \gamma$.
Monopole--nucleus bound states may exist for nuclei with a 
large gyromagnetic factor.
 \par
\ndt - {\it Energy losses of fast poles.} A fast MM 
moving  with velocity $v>10^{-2}c$ behaves like an equivalent electric
charge $(Ze)^{2}_{eq}=g^{2}\beta^{2}$. \par
\noindent - {\it Energy losses of slow monopoles ($10^{-4}<\beta<10^{-2}$)}.
 For slow particles it is important to distinguish the energy lost   
in ionization or  excitation of atoms and molecules of the medium 
(``electronic'' energy
loss) from that lost to yield kinetic energy of recoiling atoms or 
nuclei  
(``atomic'' or ``nuclear'' energy loss). Electronic energy loss predominates for 
 electrically or magnetically charged particles 
 for $\beta> 10^{-2}$. 
 The dE/dx of MMs  with $10^{-4}<\beta<10^{-3}$
is mainly due to excitations of 
atoms. 
 A monopole passing within an atom like $^{4}He_{2}$ may produce level mixings
and crossings ({\it Drell effect}) \cite{drell}.
 The effect may be used for practical 
detection  by observing the ionization caused by the energy transfer from
the excited He atoms to complex molecules with a small ionization potential 
({\it Penning effect}). \par
\noindent - {\it Energy losses at very low velocities.} 
MMs with   $v<10^{-4}c$ cannot excite atoms;
they can only lose energy in elastic collisions with atoms or with nuclei.
 The energy is released to the 
medium in the form 
of elastic vibrations and/or infra--red radiation.\par
 Fig.~2 gives a sketch of the   energy losses in liquid hydrogen  of
a $g=g_D$ MM vs its $\beta$.\par

\begin{figure}
\begin{center}
\mbox{\epsfig{file=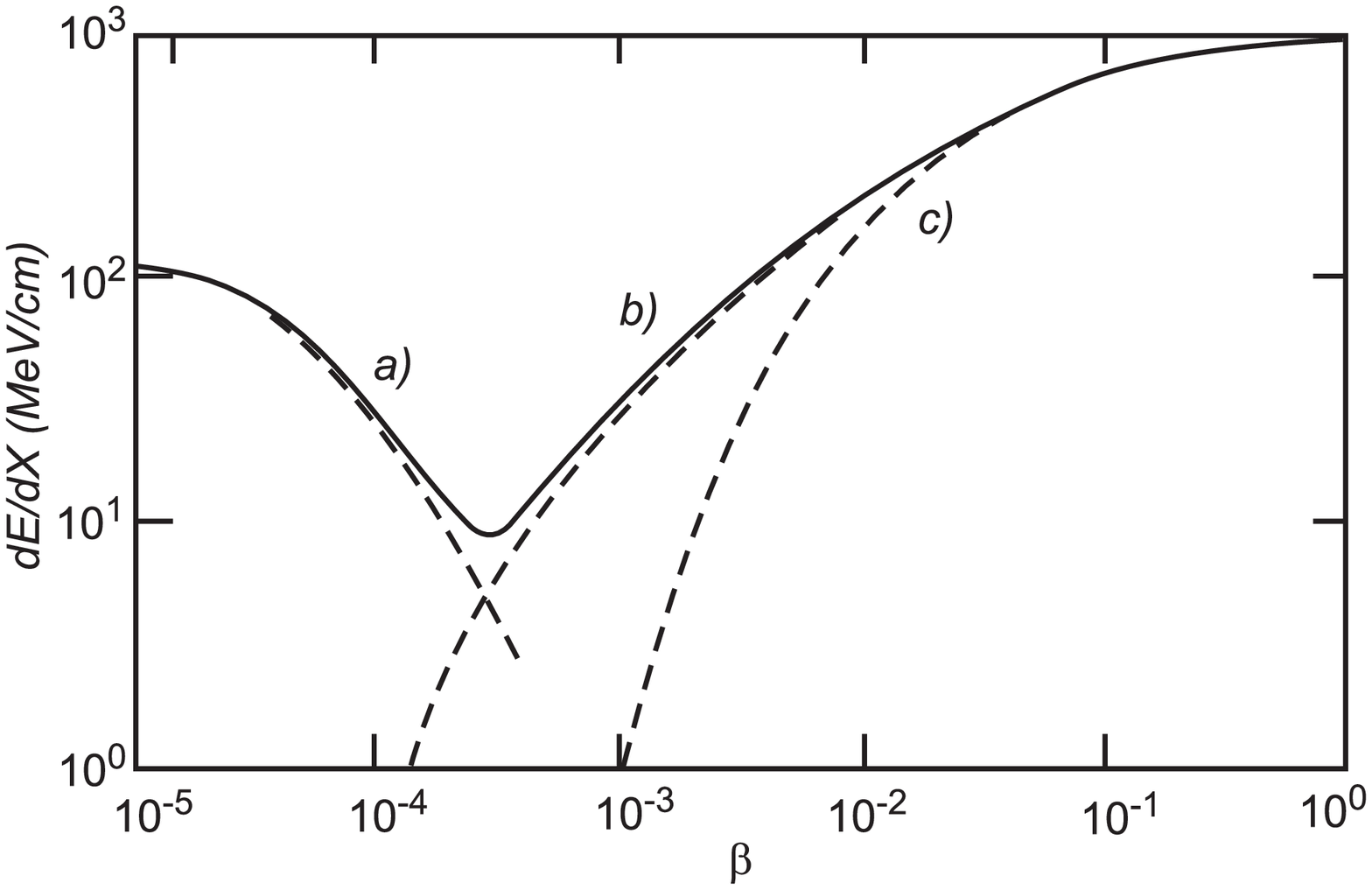,height=7cm}}
\end{center}
{\small Figure~2: The energy losses, in MeV/cm, of $g=g_D$ MMs in
liquid hydrogen as a function of ${ \beta}$. Curve a) corresponds
to elastic monopole--hydrogen atom scattering; curve b) corresponds
to interactions with level crossings; curve c) describes
the ionization energy loss.}
\end{figure}

\noindent - {\it Energy losses in superconductors.}
 If a pole
passes through a superconductor, there will be a magnetic 
flux change of $\phi_B=2\pi\hbar c/e$ (two flux quanta of superconductivity), 
 yielding $dE/dx\simeq 42$~MeV/cm, which is $\beta$-independent.\par

\noindent - {\it  Energy losses of MMs in celestial bodies.}
 For  $\beta$ $<10^{-4}$  the main energy losses in 
the earth are due to : i) pole--atom elastic scattering, 
ii) eddy current losses, 
iii) nuclear stopping power.
The earth should stop GUT  
MMs with 
$\beta\leq 10^{-4}$. From similar estimates for other celestial bodies 
one concludes that poles may be stopped if they have\\
\noindent Moon: $\beta\leq 5\times {10^{-5}}$,\quad 
Earth: $\beta \leq 10^{-4}$,\quad   
Jupiter: $\beta \leq 3\times {10^{-4}}$,\quad Sun: $\beta \leq 10^{-3}.$\par  

\section{Monopole detectors}
\par
\noindent - {\it  Superconducting induction devices.}
This method of
detection  is based only on the long--range 
electromagnetic interaction between the magnetic charge and the macroscopic 
quantum state
of a superconducting ring. 
A moving MM 
induces in the ring  an 
electromotive force and a current ($\Delta i$).
 For a  coil with N turns and inductance 
{\it L},  $ \Delta i=4\pi N ng_D/L=2\Delta i_o$, 
where $\Delta i_o$ is the current change corresponding to a change of one unit 
in the flux quantum of superconductivity (in practice 
$\Delta i \simeq 10^{-9}$A, $L\simeq$ few $\mu$H, energy 
$\simeq 4\times 10^{-17}$ erg). 
 A superconducting induction detector,
 consisting of a detection coil coupled to a SQUID (Superconducting Quantum
Interferometer Device), should be sensitive to MMs of any velocity. 

\noindent - {\it Scintillation counters.}
 Many searches have  been performed 
using excitation loss techniques. 
The light
yield from a MM in the NE110 scintillator, is shown in Fig.~3 
\cite{derkaoui1}. Curves for 
a bare $g=ng_D$ monopole  with $n$=1--9  
 are given. 
Note the presence
of a threshold at $\beta \sim 10^{-4}$, above which the light signal 
is  large compared
to that of a relativistic muon. 
The light yield in Fig.~3 shows the  saturation effect present 
in solid materials at medium velocities. For $\beta >0.1$ the light yield 
 increases because of the production of many delta rays. 
Direct measurements by Ficenec et al. \cite{ficenec} from $n-p$ 
elastic scattering in liquid scintillators proved the  
sensitivity to slow protons down to $\beta\simeq 10^{-4}$.

\begin{figure}
\begin{center}
\vspace{-1cm}
\mbox{\epsfig{file=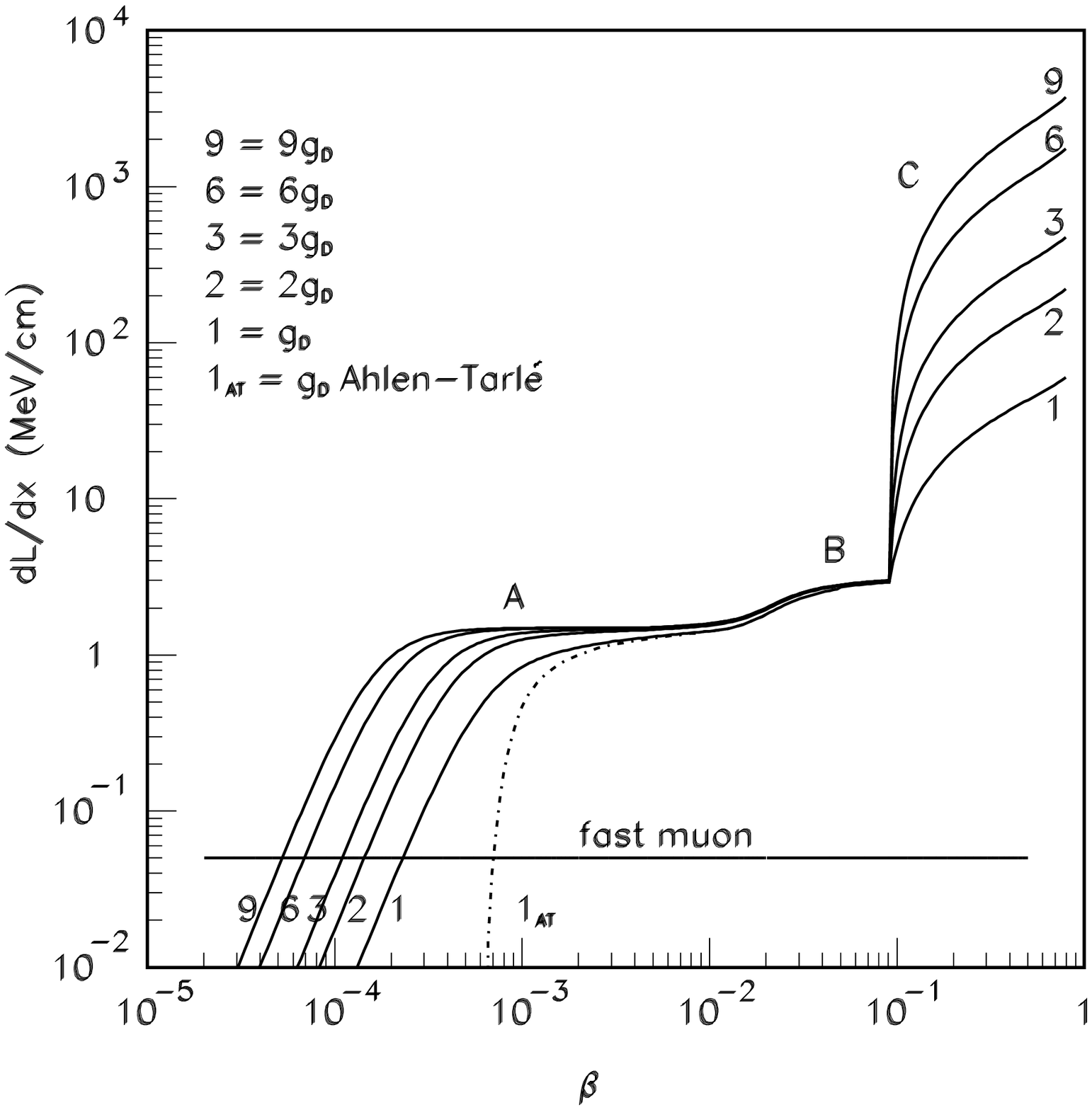,height=5.8cm}}
\vspace{4cm}
\end{center}
{\small Figure~3: Light yield of MMs in the plastic scintillator NE110
($\rho = 1.032 \;\; \mbox{g}/\mbox{cm}^{3}$) and in the MACRO liquid
scintillator ($\rho = 0.86 \;\; \mbox{g}/\mbox{cm}^{3}$),
versus $\beta$ for $g=ng_D$ magnetic charge  with $n=$1--9.}
\end{figure}

\par
\noindent - {\it Gaseous detectors.} 
Gaseous detectors of various types have been used. 
MACRO used limited streamer tubes, in  units of  8 individual tubes \cite{macro},  each equipped
with readouts for the wires and pickup strips, for two--dimensional 
localization. 
The gas   was 73\% helium and 27\% n--pentane. This
 allows  exploitation 
of the Drell and Penning effects: a magnetic monopole leaves the 
helium atoms in a metastable excited state (He*) with an excited energy of  
about 20 eV. The ionization potential of n--pentane is about 10 eV; hence, the 
Penning effect 
converts the energy of the He*  into ionization of the n--pentane 
molecule. 

\section{Searches for ``classical'' Dirac monopoles}
We shall consider as ``classical" Dirac monopoles those MMs of relatively low
mass which could be produced at accelerators.\par
\noindent - {\it Accelerator searches.} 
If MMs could be produced at high--energy accelerators, they would be 
 re\-la\-ti\-vi\-stic and  would ionize heavily. 
 Examples of direct searches are  scintillation counter searches and 
the  experiments performed 
with nuclear track detectors where data taking is integrated 
 over periods of
 months. Experiments at the Fermilab $\overline p p$ collider 
 established cross section
upper limits of $\sim 3\times 10^{-32}$~cm$^2$ for MMs with masses up to 
850 GeV. Searches at $e^{+}e^{-}$  colliders exclude masses up to 45 GeV 
\cite{gg}. 
An example of  indirect searches is the experiment at the CERN SPS; the 450 GeV
protons interacted  in a series of targets made
of  ferromagnetic tungsten powder. 
Later on the targets were placed in
front of a pulsed solenoid with a field 
$B\sim 200$ kG,  large enough to extract  and
 accelerate the MMs, to be  detected in nuclear 
emulsions and in  CR39  sheets \cite{gg}.

\ndt - {\it Multi--$\gamma$ events.} 
Five peculiar photon shower events, found in nuclear plates exposed to 
high--altitude cosmic rays, are characterized by 
an  energetic narrow cone of tens of photons, without any incident charged 
particle. The total energy in the photons is of the order of $10^{11}$ GeV. The
small 
radial spread of photons  suggests a c.m. $\gamma>10^3$. 
The energies of the photons in the overall c.m. 
system are  small, too low to have $\pi^o$ decays as 
their source. 
One possible explanation of these events could be the following: a 
high--energy $\gamma$--ray, with energy  $>10^{12}$ eV, produces in the
plate a pole--antipole pair, which then suffers bremsstrahlung and annihilation 
producing the final multi--$\gamma$ events. 
  ISR experiments, 
at $\sqrt{s}=53$ GeV, placed a  cross--section  upper--limit of 
$10^{-37}$ cm$^2$ for multi--$\gamma$ events \cite{gg}.\par

\noindent - {\it Searches in bulk matter.} 
 Classical MMs could be produced  by cosmic
rays and 
 could stop at the surface of
the earth, where they could be trapped in ferromagnetic 
 materials. It is improbable 
that GUT poles 
would stop close to the surface of the earth.
 A search 
for MMs in bulk matter used a total of 331 kg of 
material, including meteorites, schists, ferromanganese 
nodules, iron ore and other materials. The detector was a superconducting 
induction coil connected to a SQUID.
The material was passed at constant velocity through the 
magnet bore. The passage of a MM 
trapped in a sample would cause a jump in the current in the 
superconducting coil.
 From the 
absence of candidates the authors conclude that the monopole/nucleon ratio in 
the sample is $<1.2\times 10^{-29}$ at 90\% C.L.\par
Most of the searches for classical MMs performed until
1981 were not relevant to the question of the existence of very massive poles.
Ruzicka and Zrelov  summarized  all
searches for classical monopoles performed before 1980 \cite{ruzicka}.\par
\section{Cosmological and astrophysical bounds on GUT poles}
Rough, order of magnitude upper limits for a GUT monopole flux in the cosmic radiation 
were obtained on the basis of cosmological 
and astrophysical considerations. 

\ndt - {\it Limit from the mass density of the universe.} 
 This
 bound may be obtained requiring that the present MM mass density be 
smaller than the critical density $\rho_c$ of the universe. 
 For $m_M\simeq 10^{17}$ GeV one has the following 
 limit:
 $ F={n_Mc\over 4\pi}\beta<3\times  
10^{-12}h^2_0\beta~(\mbox{cm}^{-2}\mbox{s}^{-1} \mbox{sr}^{-1}).$
 It is valid for poles uniformly distributed in the universe. If poles are 
clustered in galaxies the flux limit could be few orders of magnitude larger.

\ndt - {\it Limit from the galactic magnetic field. The Parker limit.}
 The magnetic field 
in our Galaxy of $\sim 3\mu $ G is stretched in the  direction of the spiral
 arms; 
it is probably due to the non--uniform rotation of the Galaxy. This 
mechanism generates a field with a time--scale approximately equal to 
the rotation period of the Galaxy $(\tau\sim 10^8$ yr). Since MMs  
are accelerated in magnetic fields, they  gain energy, which is taken 
from the stored magnetic energy. An upper bound for the monopole flux may by  
obtained by requiring that the kinetic energy gained per unit time by 
 MMs  be less than or  
equal to  the magnetic energy generated by the dynamo 
effect. This yields the so--called Parker limit \cite{parker}. 
The original limit, 
$F<10^{-15}~\mbox{cm}^{-2}~\mbox{s}^{-1}$ sr$^{-1}$, was re--examined to take 
 account of the 
almost  
chaotic nature of the galactic magnetic field, with domain lengths of about 
$\ell\sim 1$ kpc; the limit becomes  mass 
dependent \cite{parker}. 
More recently an extended Parker bound was obtained by considering the survival of 
an early seed field \cite{adams}. The result was
$ F\leq 1.2 \times 10^{-16}(m_M/10^{17}GeV)~\mbox{cm}^{-2}~\mbox{s}^{-1}~
\mbox{sr}^{-1}.$
\par
\ndt - {\it Limit from the intergalactic field.}
Assuming the existence in the local group of galaxies of an 
intergalactic field $B_{IG}\sim 3\times 10^{-8}~G$ with a regeneration time 
$\tau_{IG}\sim 10^9~y$ and  applying the same reasoning discussed above, a 
  more 
stringent  bound is obtained;
 the
 limit is less reliable because the intergalactic field is less known.
\par
\ndt - {\it Limits  from peculiar A4 stars and from pulsars.}
Peculiar A4 stars have their magnetic fields 
$(B\sim 10^3~G)$ in the direction opposite to that expected from their 
rotation. 
 A MM with $\beta\leq 10^{-3}$ would be stopped in A4 stars;
 thus the number of MMs in the 
star would increase with time (neglecting $M\ov M$ annihilation inside the 
star). The poles could be accelerated in the magnetic field, which would 
therefore decrease with increasing time. Repeating the Parker argument, 
 one may obtain 
 strong limits, but it is not clear how good are all the 
assumptions made. 
With similar considerations applied to the superconducting core 
of neutron stars, the 
field survival of a pulsar gives an upper limit of the monopole flux in the 
neighbourhood of the pulsar. 
The limit would be particulary stringent for pulsar PSR 1937+214.

\section{Searches for supermassive GUT monopoles}

A flux of cosmic GUT supermassive magnetic monopoles may reach the earth
and may have done so for the whole life of the earth. The velocity
spectrum of these MMs  could be in the range $3 \times 10^{-5} <\beta <0.1$,
with possible peaks corresponding to the escape velocities from the earth,
the sun and the galaxy.
Searches for such MMs in the penetrating cosmic radiation 
have been performed with superconducting induction
devices whose combined limit is at the level of
$2 \times 10^{-14}~cm^{-2}~s^{-1}~sr^{-1}$, independent of $\beta$.
Several direct searches were performed above ground and underground using
scintillators, gaseous detectors and nuclear track detectors
(mainly CR39). The largest array was the Ohya one, with 
S=2000 m$^2$ of nuclear track detectors \cite{orito}. 
The most complete search was 
performed by the MACRO detector, with three different types of subdetectors and
with an acceptance of about 10,000 m$^2$sr for an isotropic flux 
\cite{mm_macro}.
 No
monopoles have been detected;  the present 90\% C.L. flux
limits are shown  in
Fig.~4 vs $\beta$ \cite{icrc01}. The limits are at the level of $2\times 10^{-16}~cm^{-2}~s^{-1}~sr^{-1}$.\par
Some indirect searches  used ancient  mica, which has a high threshold. 
 The mica experiment 
scenario assumes that a bare monopole passing through the earth captures an 
aluminium nucleus and drags it through subterranean mica causing a trail of 
lattice defects. As long as the mica is not reheated, the damage trail will
survive. The  mica pieces analyzed are small 
($13.5$ and $18$ cm$^2$), but should have been recording tracks 
since they cooled,  
about $4\div9\times 10^8$ years ago. The 
  upper--limit fluxes are at the level of 
$10^{-17} ~\mbox{cm}^{-2}~ \mbox{s}^{-1} $sr$^{-1}$ for $10^{-4}<\beta<10^{-3}$
\cite{price}.
There are many reasons why these indirect experiments 
 might not be 
sensitive. For example, if MMs have a positive electric 
charge  or have protons attached then Coulomb repulsion could 
prevent capture of heavy nuclei.\par

\begin{figure}
\vspace{-2cm}
\begin{center}
\mbox{ \epsfysize=10.5cm
            \epsffile{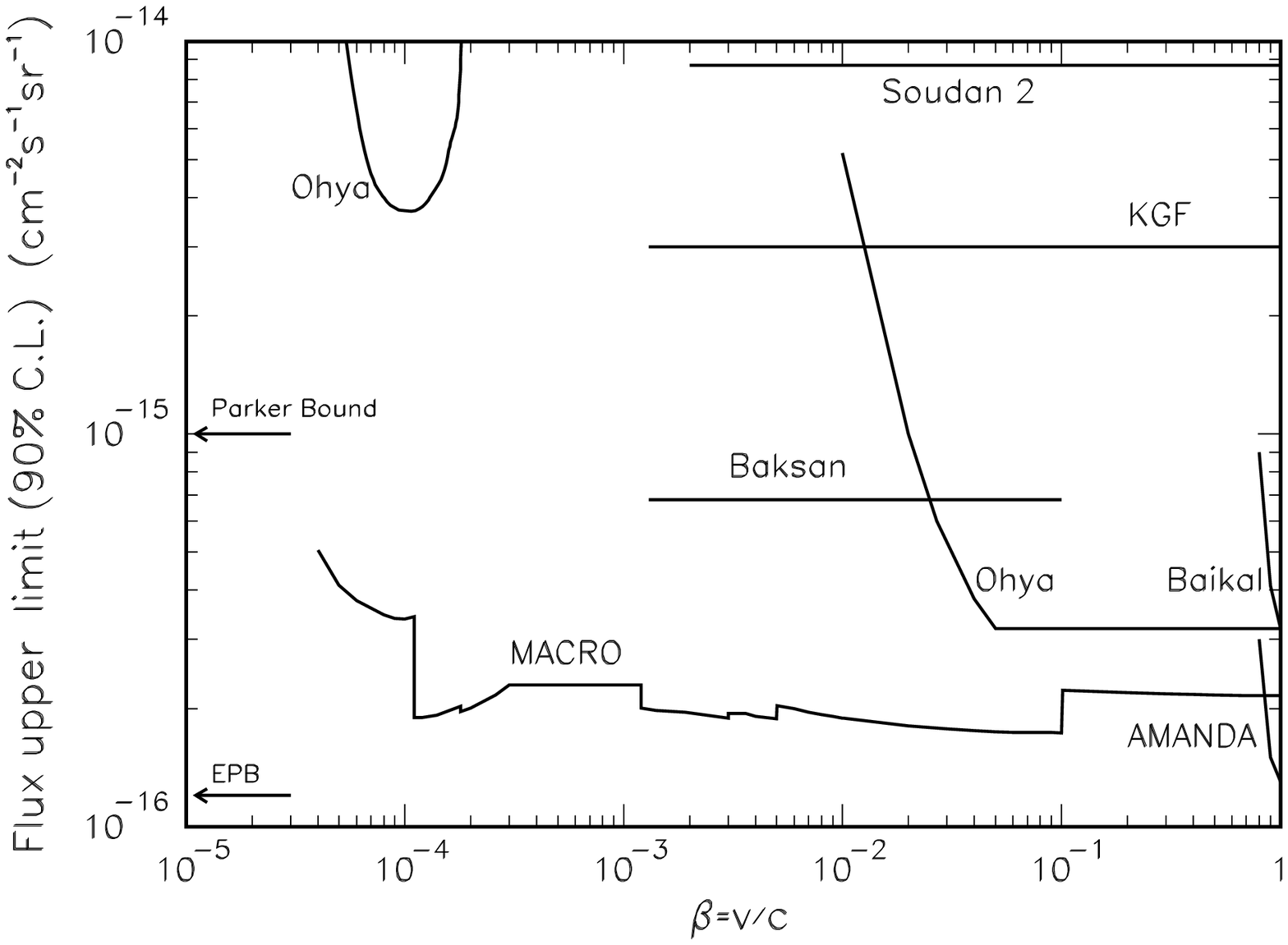}}
\end{center}
{\small Figure~4:
Compilation at 90\% C.L. of direct experimental upper limits on an isotropic
MM flux reaching detectors at the earth surface or
underground.}
\end{figure}
 
\section{Intermediate mass magnetic monopoles}
Relativistic magnetic monopoles with intermediate masses, $10^5~<m_M<~10^{12}$ 
GeV, could
be present in the cosmic radiation. Detectors at the earth surface 
would be capable to detect MMs
coming from above if they have masses larger than $\sim 10^5-10^6$ GeV,
see Fig.~5 \cite{derkaoui2}; lower
mass monopoles may be searched for with detectors located at high mountain
altitudes, or even higher, in balloons and in satellites. Few experimental
results are available \cite{nakamura}. The limit from the AMANDA experiment 
under ice 
at the south pole is shown in Fig.~4 \cite{amanda}.

The SLIM experiment is searching for IMMs
with nuclear track detectors at the Chacaltaya
high altitude lab (5230 m above sea level) \cite{slim}.

\begin{figure}
\begin{center}
        \mbox{ 
\epsfysize=8.4cm
\epsffile{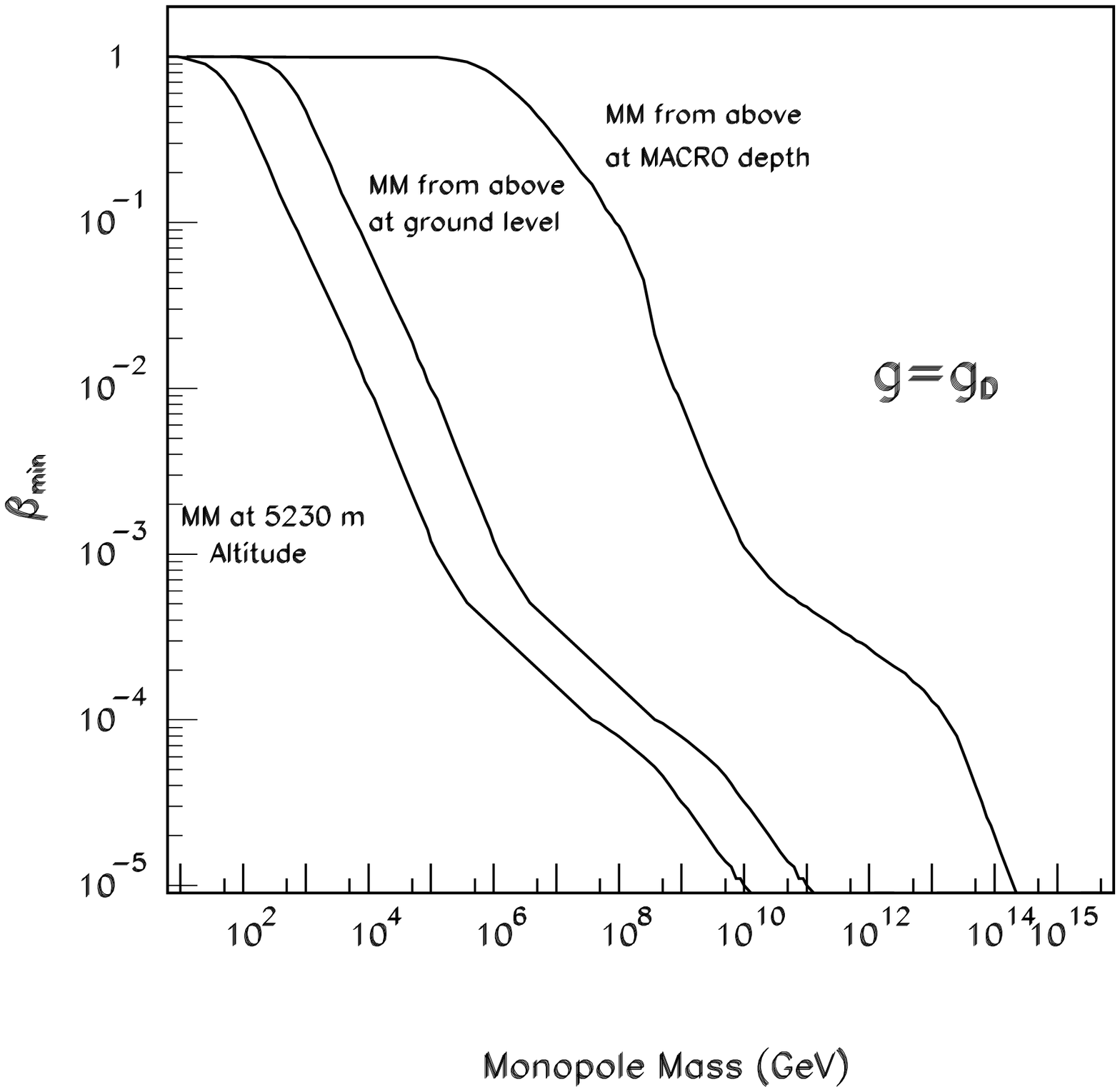}}
\end{center}
{\small Figure~5:
Accessible region in the plane (mass, $\beta$) for monopoles with
magnetic charge $g= g_D$ from above for an experiment at an altitude of
5230 m, at sea level and
for an underground detector at the Gran Sasso Lab (at an average depth of 3700
hg/cm$^2$).}
\end{figure}

\section{Monopole catalysis of proton decay}

A GUT pole may catalyze  proton decay, $ p+M\to M+e^+ +\pi^0$. 
The cross--section 
could be comparable with that of ordinary
strong interactions, if the 
MM core is surrounded by a 
fermion--antifermion condensate (Fig.~1), with some
 $\Delta B\ne 0$ terms extending up to the confinement region. 
 Thus  MMs may capture a proton or a nucleus and 
lead to the catalysis reaction. 
For spin 1/2 nuclei, like aluminium, 
there should be an enhancement in the cross section over that for free protons.
 Instead for spin--0 nuclei there should be a  $\beta$--dependent
 suppression.
 For oxygen the suppression factor 
could be of the order of $10^{-2}$ at $\beta=10^{-3}$, $\simeq 10^{-5}$ at 
$\beta=10^{-4}$. \par
If the $\Delta B\ne 0$ cross--section for MM catalysis of proton decay 
were large, then a monopole would trigger a chain of baryon ``decays'' along 
its passage through a large detector \cite{icrc01}.\par
It should be noted that if MMs have a large catalysis cross--section then 
the monopole--proton 
composites could be unstable.
\par
\noindent - {\it Astrophysical limits from monopole catalysis of nucleon 
decay.} 
The number of MMs inside a star or a planet should  increase with
time, due to a constant capture rate and a  small pole--antipole
annihilation rate. The catalysis of nucleon decay by MMs could 
be another source of energy for these astrophysical bodies.
The catalysis argument, applied to the protons of our sun, leads to the 
possibility that
the sun could emit high energy electron neutrinos.
 The $\nu_e$'s 
could be detected through their elastic scattering 
on electrons. The Kamiokande experiment quoted the limit 
$F<8\times 10^{-10}\beta^2$ if the monopole catalysis 
cross--section is 1 mb.
From such limits  one could place 
a limit on the number of poles in the sun,  of less 
than 1 pole per $10^{12}$ g of solar material \cite{gg}.\par

A speculative upper bound on the total number of MMs	
 present inside the earth can be made 
assuming that the energy released by MM 
catalysis of nucleon decay in the earth does not exceed the surface heat 
flow. 

\section{Nuclearites}

{\it Strangelets, Strange Quark Matter (SQM)} should consist of aggregates
of $u,~d$ and $s$ quarks in almost
equal proportions (the number of $s$ quarks should be lower than the number
of $u$ or $d$ quarks; thus the SQM should have a positive charge \cite{witten}. 
The SQM should be a colour singlet; thus
it should have
only integer electric charge. The overall neutrality of SQM
is ensured by
an electron cloud which surrounds it, forming
a sort of atom. (We shall use the word nuclearite to denote the 
core+electron system).\par
 Strangelets could have been produced shortly after the Big Bang and may
have survived as remnants; they could
also appear in violent astrophysical
processes, such as neutron star collisions.
Nuclearites should have a constant density \cite{derujula2},
 $\rho_N = M_N/V_N \simeq 3.5 \times 10^{14}$ g cm$^{-3}$, somewhat
larger than that of atomic nuclei, and
they should be stable for all baryon
numbers in the range between ordinary heavy nuclei and neutron stars
($A \sim 10^{57}$) \cite{derujula2}. Nuclearites could contribute to the cold
dark matter.   \par
The relation between mass and size of nuclearites is illustrated in Fig.~6. \par

\begin{figure}[h]
\begin{center}
\mbox{\hspace{-0.5cm}
\epsfig{file=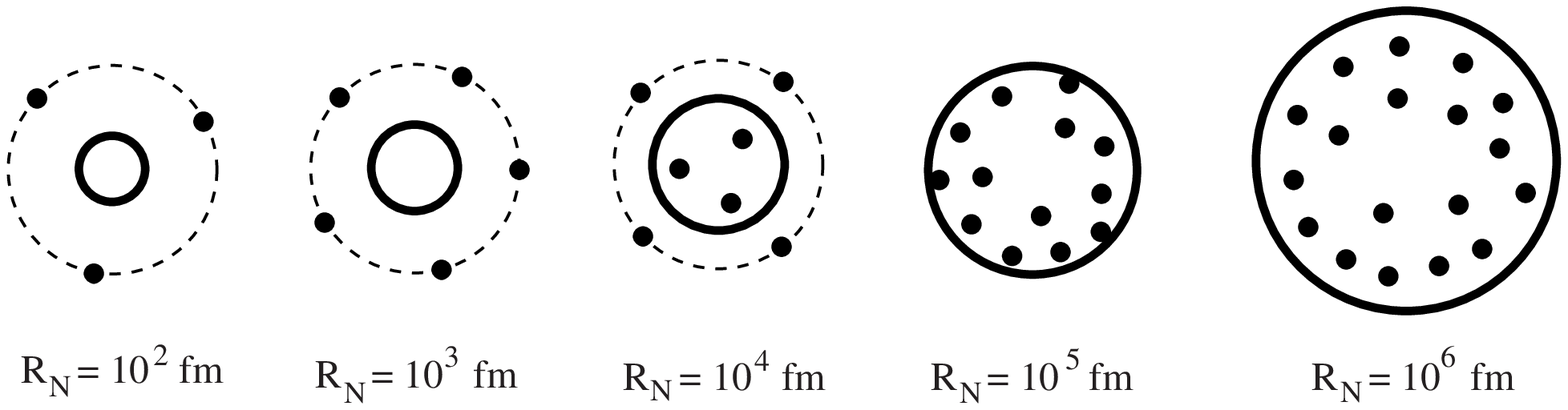,height=3.35cm}}
\vspace{0.5cm}
\end{center}
{\small Figure~6: Dimensions of the quark bag ($R_N$) and of
the core+electrons system (nuclearite).
 The radii presented here (in a logarithmic scale) refer to the nuclearite
quark bag core. For nuclearite
masses smaller than $10^{9}$ GeV/c$^2$, the entire electron cloud is outside the
quark bag and the core+electrons system  has a global size
of approximately $10^{5}$ fm = 1 \AA; for $10^{9} < M_N < 10^{15}$
GeV/c$^2$ the electrons are partially inside the core; for $M_N > 10^{15}$
GeV/c$^2$ all electrons are inside the core. The black dots indicate the
electrons, the quark bag border is indicated by thick solid lines; the
border of the core+electronic cloud system for relatively small masses
is indicated by the dashed lines.}
\end{figure}

The main energy loss mechanism for low velocity nuclearites passing through matter is that
of atomic collisions. While traversing a medium the nuclearites should
displace the
matter in their path by elastic or quasi-elastic collisions with the ambient
atoms \cite{derujula2}.
The energy loss rate is large; therefore nuclearites should be easily 
detectable by detectors (like scintillators and CR39 nuclear track detectors)
 used for MM searches. \par
Nuclearites are expected to have typical galactic velocities, $\beta 
\sim 10^{-3}$. For such 
velocities, nuclearites with masses larger than 0.1 g could traverse the earth.
Most nuclearite searches were obtained as byproducts of superheavy magnetic
monopole searches. The cosmic ray flux limits are therefore similar to
those obtained for MMs.\par
The most relevant direct flux upper limits for nuclearites come from three
 large area experiments: the first two use  CR39 nuclear track detectors;
one experiment was
performed at mountain altitude \cite{nakamura}, the second at a depth of
10$^4 g~cm^{-2}$ in the Ohya
mines \cite{orito}; the third experiment was  MACRO which used liquid
scintillators besides nuclear track detectors \cite{macro2}. 
A fourth experiment (SLIM) 
is deployed at high altitudes \cite{slim}. Indirect experiments
using old mica
samples could yield  the lowest flux limits, but they are affected by inherent
systematic uncertainties [25].
Some exotic cosmic ray events were interpreted as due to incident nuclearites,
for example the ``Centauro" events and the anomalous massive particles [33].
The
interpretation of those possible signals are not unique and the used
detectors are not redundant enough to reach a conclusion.\par
In Fig.~7 is presented a compilation of limits for a flux of downgoing
nuclearites compared with the dark matter limit, assuming a velocity at
ground level  $\beta = v/c =2 \times 10^{-3}$. This speed corresponds to
nuclearites of galactic or extragalactic
origin. In the figure the MACRO limit was extended above the dark matter bound,
in order to show the transition to an isotropic flux for nuclearite masses
larger than 0.1 g ($\sim 10^{23}$ GeV).

\begin{figure}[h]
\begin{center}
\mbox{\epsfig{file=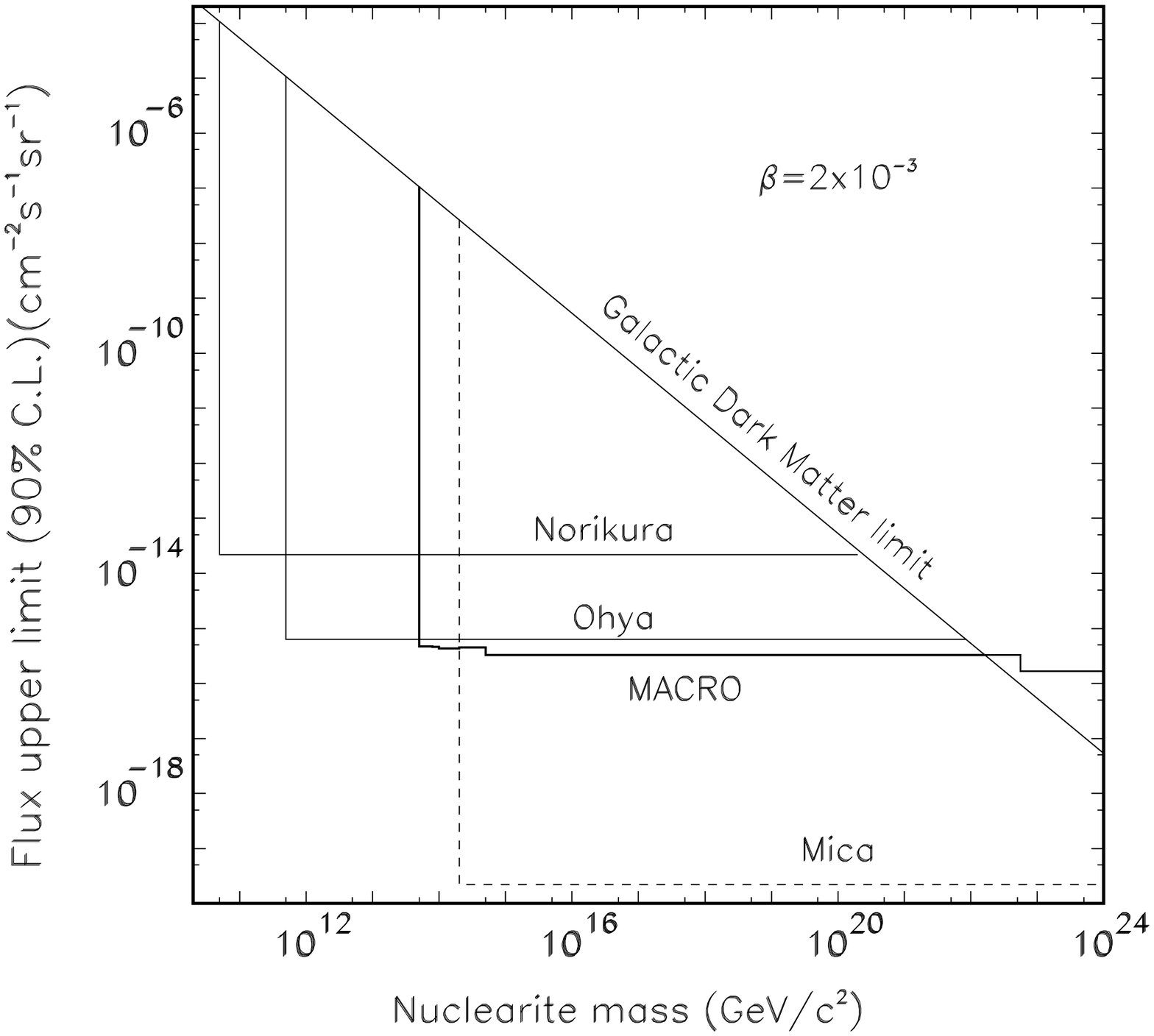,height=9.2cm}}
\vspace{-0.5cm}
\end{center}
{\small Figure~7: 90\% C.L. 
flux upper limits versus mass for nuclearites with
$\beta=2\times~10^{-3}$ at ground level. These nuclearites
could have galactic or extragalactic origin. The limits are from
  MACRO [16],  from Refs.~[19] (``Nakamura''), [15] (``Orito'') 
and the indirect
Mica limits of Ref.~[17].}
\end{figure}

\section{Q-balls}

Q-balls  should be aggregates of 
squarks $\tilde q$, sleptons $~\tilde l$ and 
Higgs fields \cite{coleman,kusenko1}.  The scalar condensate inside 
a Q-ball core
has a global baryon number $Q$ (and may be also a lepton number). 
Protons, neutrons and may be electrons
could be absorbed in the condensate. \par
There could exist neutral and charged Q-balls:
 Supersymmetric Electrically Neutral Solitons (SENS), which
do not have a net electric charge, are generally massive and may 
catalyse proton decay. SENS may obtain an integer positive electric charge
absorbing a proton in their interactions with matter yielding 
 SECS (Supersymmetric Electrically Charged Solitons), which
have a core 
electric charge,  have generally lower masses and the Coulomb
barrier could prevent the capture of nuclei. SECS have only integer
charges because they are colour singlets. Some
Q-balls which have sleptons in the condensate can  also absorb
 electrons. The squarks $\tilde q$ inside the scalar 
potential bag should have essentially zero masses. \par
A SENS which enters the earth atmosphere could absorb a
nucleus of nitrogen which would give it the positive charge of $+7$ (SECS with
$Z= +7$). Other
nuclear absorptions are prevented by Coulomb repulsion. If the  
Q-ball can absorb electrons at the same rate as protons, the positive charge
of the absorbed nucleus may be neutralized by the charge of absorbed 
electrons.
 If, instead, the absorption of electrons is slow or 
impossible, the Q-ball carries a positive electric charge after the capture
of the first nucleus in the atmosphere. \par
The Q-balls could be possible cold dark matter 
candidates. 
Flux limits on Q-balls may come
from  astrophysical dark matter limits.
 SECS
with $\beta \simeq 10^{-3}$ and $M_Q < 10^{13}$ GeV/c$^2$ could reach 
an underground detector from above, SENS also from below 
\cite{kusenko2,kusenko3}.
 SENS may be detected by their almost continuous emission of charged pions 
(energy loss of about 100 GeV g$^{-1}$cm$^{2}$), while SECS may be detected by
their large energy losses yielding light in scintillators, and 
possibly ionization.

\section{Conclusions. Outlook}
\vspace{-0.5cm}$    $\par
1. Direct and indirect searches for classical Dirac monopoles have placed mass 
limits at the level of $m_M > 850$ GeV. Future improvements could come from
experiments at the LHC.\par
2. Many searches have been performed for superheavy GUT monopoles in the 
penetrating cosmic radiation. The flux limits are at the level of 
$ \Phi \le 2 \times 10^{-16} cm^{-2}s^{-1}sr^{-1}$ for $\beta \ge 3
\times 10^{-5}$.
It would be difficult to do much better since one would require detectors of
considerably larger areas. Or one has to devise new techniques.\par
3. Present limits on Intermediate Mass Monopoles are relatively poor. 
Experiments at high altitudes and at neutrino telescopes may be able to improve the
situation.\par
4. For nuclearites with typical galactic velocities one may repeat the 
considerations made in points 2 and 3. For them the searches at high altitude labs are
very important.\par
5. For Q-balls the situation is less clear, though some considerations similar
to those of point 4 can be made.

\section{Acknowledgements}We would like to acknowledge the cooperation of
 many colleagues of the MACRO collaboration, in particular F. Cei, M.Cozzi,
I. De Mitri,   
M. Giorgini, F. Guarino, G. Mandrioli, M. Ouchrif, V. Popa, P. Serra, M. Spurio, and  
others.

\footnote{A bibliography compilation on magnetic monopoles updated to the end
of 1999 can be found in Ref.~\cite{biblio}.}

\end{document}